\documentclass[jcp,twocolumn,showpacs,preprintnumbers,superscriptaddress]{revtex4-1}
\usepackage[bookmarks, colorlinks=true, plainpages = true,linkcolor = blue, citecolor = blue, urlcolor = blue, filecolor =blue]{hyperref}

\usepackage{amsfonts}
\usepackage{graphicx}
\usepackage{amsmath}
\usepackage{amssymb}
\usepackage{latexsym}
\usepackage{psfrag}
\usepackage{dcolumn}
\usepackage{datetime}
\usepackage{multirow}
\usepackage{subfigure}

\begin{document}
\title{Scanning tunneling spectroscopy and Dirac point resonances due to a single Co adatom on gated graphene}
\author{Alireza Saffarzadeh}
\affiliation{Department of Physics, Payame Noor University, P.O.
Box 19395-3697 Tehran, Iran} \affiliation{Department of Physics,
Simon Fraser University, Burnaby, British Columbia, Canada V5A
1S6}
\author{George Kirczenow}
\affiliation{Department of Physics, Simon Fraser University,
Burnaby, British Columbia, Canada V5A 1S6}
\date{\today}

\begin{abstract}
Based on the standard tight-binding model of the graphene
$\pi$-band electronic structure, the extended H\"{u}ckel model for
the adsorbate and graphene carbon atoms, and spin splittings
estimated from density functional theory (DFT), the Dirac point
resonances due to a single cobalt atom on graphene are studied.
The relaxed geometry of the magnetic adsorbate and the graphene is
calculated using DFT. The system shows strong spin polarization in
the vicinity of the graphene Dirac point energy for all values of
the gate voltage, due to the spin-splitting of Co $3d$ orbitals.
We also model the differential conductance spectra for this system
that have been measured in the scanning tunneling microscopy (STM)
experiments of Brar {\em et al.} [Nat. Phys. {\bf 7}, 43 (2011)].
We interpret the experimentally observed behavior of the S-peak in
the STM differential conductance spectrum as evidence of tunneling
between the STM tip and a cobalt-induced Dirac point resonant
state of the graphene, via a Co $3d$ orbital. The cobalt
ionization state which is determined by the energy position of the
resonance can be tuned by gate voltage, similar to that seen in
the experiment.
\end{abstract}
\maketitle

\section{Introduction}
Graphene, a single atomic layer of graphite, has attracted a great
amount of attention because of its fundamental interest and
promising applications \cite{Novoselov1,Zhang1,Geim1,Geim2}. One
of the remarkable properties of pristine graphene is the linear
dispersion relation of its electronic spectrum near the Dirac
point \cite{Wallace,Neto}. It has been demonstrated theoretically
that chemical adatoms, whether magnetic
\cite{Mao,Chan0,Wehling5,Wehling6,
Can,Jacob,Chan,Rappoport,Wehling8,Liu} or nonmagnetic
\cite{Liu,Skrypnyk1,Pereira1,Wehling1,
Skrypnyk2,Robinson,Pereira2,Basko,
Wehling2,Wehling3,Pershoguba,Wehling4,Skrypnyk3,George1}, adsorbed
on graphene can induce resonant states in the vicinity of the
Dirac point energy, and these states may give rise to strong
scattering of electrons in the graphene. However, to our
knowledge, there has as yet been no direct experimental
observation of these Dirac point resonances. The presence of
chemical adsorbates, atomic or molecular, on graphene strongly
affects electronic properties and makes graphene-based devices
suitable for chemical sensing
\cite{Schedin,Wehling7,Leenaerts,Romero,Zhang,George1,
Ao2010,Song12}.

Furthermore, the chemical potential of graphene electrons can be easily
tuned by a gate voltage, and thus, the adsorbate ionization state
can be externally controlled \cite{Jacob,Chan}. Using single and
few-layer graphene films on top of a SiO$_2$ substrate, it has
been demonstrated that a graphene field-effect transistor can
switch between two dimensional electron and hole gases by changing
the gate voltage \cite{Novoselov2}. It has been shown that, using
an electrochemically top-gated graphene transistor, even much
higher electron and hole doping can be induced than by standard
SiO$_2$ back-gating \cite{Das}. On the other hand, an insulating
state with large suppression of conductivity can be created in
bilayer graphene field-effect transistors by using a top gate in
addition to the global back gate \cite{Oostinga}. When an electric
field was applied perpendicular to the bilayer, a gap of less than 10
meV at low temperatures \cite{Oostinga} and up to 0.25 eV at room
temperature \cite{Zhang2} opened in the bilayer graphene, the size
of the gap being tunable by the electric field.

On the other hand, according to density functional theory (DFT)
calculations, cobalt adatoms on graphene show the highest magnetic
anisotropy energy relative to other $3d$ transition metals
adsorbed on graphene \cite{Xiao,Sargol}. Thus, the Co-graphene
system has potential applications in magnetic data storage.
Recently, scanning tunneling microscopy (STM) and spectroscopy
measurements demonstrated that individual Co atoms deposited onto
back-gated graphene devices can be controllably ionized by
application of a gate voltage or STM tip bias voltage \cite{Brar}.
Several different features were observed in the gate-dependence of
the measured differential conductance. Inelastic electron
tunneling and hybrid electronic-vibrational states were suggested
as possible explanations for some of these features. Therefore,
the effects of inelastic scattering of electrons should be
included in any theoretical study whose objective is to reproduce
these features of the calculated inelastic tunneling spectra
\cite{Stipe,Qiu,Wehlinginelastic,Firuz}. A different feature is related to a
sharp peak in the spectrum which moves in the opposite direction
in energy than the graphene Dirac point energy in response to
changes in the gate voltage.

In a previous theoretical study, using DFT electronic structure
calculations in a generalized gradient approximation for periodic
systems, Jacob and Kotliar \cite{Jacob} showed that a single Co
atom most likely adsorbs at the hollow site of a clean graphene
sheet. They found that the coupling of the Co $3d$ levels to the
graphene substrate and the dynamic correlations are strongly
dependent on the orbital symmetry, temperature, distance of the Co
atom from the graphene sheet, and gate voltage. More recently,
Chan {\em et al.} \cite{Chan} investigated the gated Co-graphene
system in a supercell structure with periodic boundary conditions
using the local density approximation (LDA)+\textit{U} method,
with several values of the \textit{U} parameter. The projected
density of states was calculated for different \textit{U} values
and back gating treated qualitatively by LDA+\textit{U}
calculations for different integer charge states of the sample per
unit cell. An ionization effect for the Co adatom was found as the
charge state of the whole system was varied. No calculated STM
current-voltage characteristics were reported, so the calculated
local density of states (LDOS) could not be compared with STM
data. Hence, the true LDOS of the Co-graphene system remains
uncertain. Note that experimental and theoretical studies of the
interaction between Co and graphene have shown charge-transfer to
occur between the graphene and Co, resulting in the graphene being
electron doped \cite{Wang,Vanin,Eom}.

The aim of this paper is to reproduce some resonance features seen
in the STM experiment by means of DFT calculations of the relaxed
geometry and the extended H\"{u}ckel model of quantum chemistry
\cite{George2,Huckel1,Huckel2} for the electronic structure of Co
adatom on graphene with DFT-based spin splitting parameters. To
this end, we study resonant scattering of spin-up and spin-down
electrons by the single Co atom adsorbed on gated graphene for
electron energies close to the Dirac point. The electronic density
of states for the Co-graphene system is calculated and the
dependence of the resonant states on the gate voltage is
discussed. In addition, we investigate the differential
conductance of electron tunneling through the Co adatom as a
function of sample bias and gate voltage which has not been
addressed in the previous theories. Our transport calculations
show that the experimentally observed behavior of the previously
unidentified S-peak in the STM differential conductance spectrum
\cite{Brar} is consistent with this feature being due to tunneling
between the STM tip and a cobalt-induced Dirac point resonant
state of the graphene, via a Co $3d$ orbital. In this study, we do
not consider the Kondo effect of Co on graphene
\cite{Wehling5,Wehling6,Jacob}, because this is a many-body effect
and outside of the scope of the present study.

The paper is organized as follows. In Sec. II, we present the
model Hamiltonian of the infinite 2D graphene sheet with a single
cobalt adatom. We show that the tight-binding Hamiltonian for the
system can be transformed into an effective graphene Hamiltonian
that includes the effect of the Co adatom. In Sec. III, we derive
an exact relation for the graphene Green's function matrix
elements in the presence of the adatom using $T$-matrix theory.
The results for the spin- and gate-dependent $T$-matrix and the
electronic density of states will be presented in Sec. IV, where
the adsorbate-induced resonance states are discussed. The effect
of local rehybridization of the graphene from the $sp^2$ to $sp^3$
electronic structure on the Dirac point resonances, that occurs
when the Co atom bonds covalently to the graphene, is also
examined. In Sec. V, we present a formalism suitable for studies
of the gate-induced ionization of a single Co adatom on graphene,
seen in the experiment. We calculate the gate dependence of the
differential conductance by means of the Landauer-B\"{u}ttiker
formalism based on the Green's function technique, and the
extended H\"{u}ckel theory and DFT-based spin splitting parameters
and compare the results with the STM measurements of Brar {\em et
al.}\cite{Brar}. Finally, in Sec. VI we conclude this work with a
general discussion of the results.

\section{The Model}
Our calculations are performed using \textit{ab initio} geometry
relaxation based on density functional theory for a Co adsorbate
on the honeycomb graphene lattice using the Gaussian 09 software
package together with the HSEh1PBE hybrid function and the
6-311G(d) Gaussian basis set \cite{George2,Huckel1,Huckel2}. Such
a relaxed geometry is expected to be accurate because the density
functional theory on which this relaxation is based, has already
been well optimized for carrying out accurate ground state total
energy calculations. We found that the lowest-energy binding site
is the H-site in which the adsorbate is positioned at height
$h_0=1.68$ {\AA} above the middle of a hexagon of the graphene, as
shown in Fig. 1. The reason is that, the H-site has the largest
number of neighboring C atoms, lowering the adsorption energy. The
structure studied was a graphene disk consisting of 54 carbon
atoms passivated at the edges with 18 hydrogen atoms and the Co
atom being bonded to the graphene at the center of the disk. All
the atoms were allowed to relax freely. Although a finite size
cluster of carbon atoms was used in the calculations, the bond
lengths between the Co atom and its neighboring C atoms were well
converged which indicates that 54 C atoms is large enough to model
the graphene plane. The final relaxed structure obtained in this
way is shown in Fig. 1. From the figure, it is clear that, there
is no distortion out of plane of the graphene structure due to the
Co adsorbate. This is in agreement with the previous studies of
Co-graphene systems in periodic supercell structures
\cite{Mao,Chan}. In addition, we found in-plane distortion of 0.01
{\AA} in bond length between carbon atoms of central hexagon
relative to their nearest neighbor carbon atoms in the graphene
plane. The existence of an in-plane distortion equal to 0.023
{\AA} in the periodic Co-graphene supercell has also been reported
by Liu {\em et al.} \cite{Liu}.
\begin{figure}
\centerline{\includegraphics[width=1.0\linewidth]{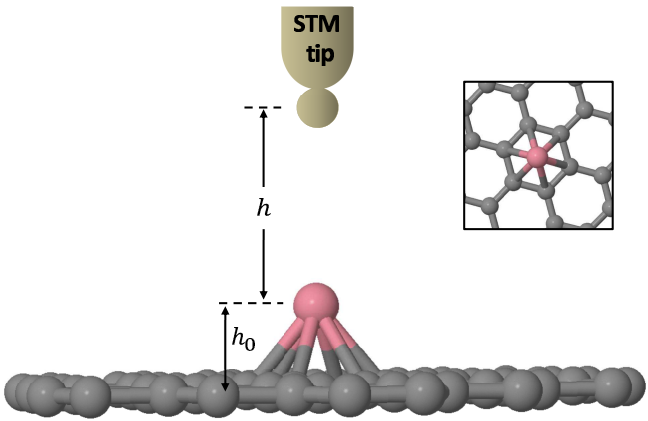}}
\caption{(Color online) Side view of the relaxed geometry for a
single adsorbed Co atom on graphene. The optimal height $h_0=1.68$
{\AA} is the perpendicular distance between the Co adsorbate and
graphene plane, while $h=4$ {\AA} is the distance between the
adsorbate and STM tip which is directly above the Co atom. The
inset shows the top view of the Co-graphene system. The hydrogen
atoms at the edges of graphene are not shown here.}
\end{figure}

We describe the present system by a tight-binding model
Hamiltonian derived from extended H\"{u}ckel theory in a basis of
\textit{extended molecular orbitals} (EMOs) which are linear
combinations of the atomic valence orbitals of the adatom and the
2s, 2$p_x$, and 2$p_y$ orbitals of the graphene carbon atoms. We
refer the reader to Ref. \cite{George1} for a detailed explanation
regarding tight-binding models for adsorbates on graphene derived
from extended H\"{u}ckel theory and to
Refs. \onlinecite{Kienle06, Raza08, Graphenebook}
for further discussions
of the application of extended H\"{u}ckel theory to graphene-based
systems and carbon nanotubes. Accordingly, the EMOs together
with the 2$p_z$ orbitals of the graphene carbon atoms form the
basis set for our total Hamiltonian of the Co-graphene system that
can be written in the form
\begin{equation}\label{01}
H=H_0+\sum_{\alpha\sigma}\epsilon_{\alpha\sigma}
d_{\alpha\sigma}^\dag d_{\alpha\sigma}+
\sum_{\alpha,\sigma,n}\gamma_{\alpha
n,\sigma}(d_{\alpha\sigma}^\dag a_{n\sigma}+\mathrm{h.c.})
\end{equation}
where the first term corresponds to the Hamiltonian of clean
graphene in the basis of 2$p_z$ carbon orbitals, $a_{n\sigma}$ is
the annihilation operator for an electron with spin $\sigma$ in
the 2$p_z$ orbital $\phi_{n\sigma}$ of carbon atom $n$,
$d_{\alpha\sigma}^\dag$ is the creation operator for an electron
in an EMO $\psi_{\alpha\sigma}$ of the Co adatom and
$\epsilon_{\alpha\sigma}$ is the corresponding energy eigenvalue.
$\gamma_{\alpha n,\sigma}$ is the spin-dependent matrix element of
the extended H\"{u}ckel Hamiltonian between the 2$p_z$ orbital of
carbon atom $n$ and the EMO $\alpha$. We note that, since the
basis set used in the extended H\"{u}ckel theory is nonorthogonal,
it is possible for the overlap $s_{\alpha
n,\sigma}=\langle\psi_{\alpha\sigma}|\phi_{n\sigma}\rangle$ between the
$2p_z$ orbital $\phi_{n\sigma}$ of carbon atom $n$ and EMO
$\psi_{\alpha\sigma}$ to be non-zero. Because of this, we replace
$\gamma_{\alpha n,\sigma}$ in the Hamiltonian by $\gamma_{\alpha
n,\sigma}-\epsilon s_{\alpha n,\sigma}$ where $\epsilon$ is the electron
energy \cite{Emberly1,Emberly2}. Henceforth we will omit the spin
index $\sigma$ in the overlap parameter and show it as $s_{\alpha n}$ for simplicity.

The unperturbed Hamiltonian of two dimensional graphene, $H_0$,
which can be modeled on a honeycomb lattice of two nonequivalent
sites (that is, $A$ and $B$) per unit cell is expressed as
\cite{Neto}
\begin{equation}\label{3}
H_0=-t\sum_{\langle i,j\rangle}\,[a_i^\dag a_j+\mathrm{h.c.}]
\end{equation}
and describes the motion of carriers between 2$p_z$ carbon
orbitals of graphene with a nearest-neighbor hopping parameter
$t$. We neglect the small changes in $t$ due to the above
mentioned in-plane distortion in the graphene geometry, induced by
the Co atom. We note that although Eq. (\ref{3}) does not include
the overlaps between the nearest neighbor $2p_z$ orbitals of the
carbon atoms {\em explicitly}, it is a widely used form of the
Hamiltonian that has been optimized to provide an accurate
description of the graphene $\pi$-band states around the Dirac
point.\cite{Neto} We adopt it in the present work for that reason.
Furthermore, since it is assumed that in clean graphene the
spectrum is spin-degenerate, for simplicity we have suppressed the
spin indices in Eq. (\ref{3}). Here $a_i^\dag$ is the creation
operator for an electron in 2$p_z$ carbon orbital $i$. The
Hamiltonian for graphene is diagonal when written in the basis
$|\mathbf{k},c(v)\rangle$ defined as $
|\mathbf{k},c\,(v)\rangle=(|\mathbf{k},A\rangle\pm
e^{-i\phi(\mathbf{k})} |\mathbf{k},B\rangle)/\sqrt{2}$ where $A$
and $B$ stand for $A$ and $B$ sublattices and
$|\mathbf{k},A(B)\rangle=\frac{1}{\sqrt{N_{A(B)}}}\sum_{\mathbf{R}_{A(B)}}
e^{i\mathbf{k}\cdot\mathbf{R}_{A(B)}}|\mathbf{R}_{A(B)}\rangle$.
Here, the $c\,(v)$ sign denotes the conduction (valence) band,
$|\mathbf{R}_{A(B)}\rangle$ represents an atomic orbital centered
at a lattice site $A(B)$ and $N_{A(B)}$ is the total number of
$A(B)$-lattice points. The phase $\phi(\mathbf{k})$ is the polar
angle for momentum $\mathbf{k}$ and is defined as
$\phi(\mathbf{k})=\mathrm{Im}(\ln f(\mathbf{k}))$ where
$f(\mathbf{k})=e^{ik_xa/\sqrt{3}}+2e^{-ik_xa/2\sqrt{3}}\cos(k_ya/2)$.
The corresponding eigenvalues in terms of $f(\mathbf{k})$ can be
expressed as $\varepsilon_{c\,(v)}=\pm t\sqrt{|f(\mathbf{k})|^2}$
\cite{Wallace,Neto}.

The carbon $2s$, $2p_x$, and $2p_y$ orbitals in graphene combine
to form in-plane $\sigma$ (bonding) and $\sigma^*$ (anti-bonding)
orbitals. In this study the $2s$, $2p_x$, and $2p_y$ orbitals of
the graphene were shifted down in energy relative to their
energies in extended H\"{u}ckel model so as to yield an energy gap
between the graphene $\sigma$ and $\pi$ bands in agreement with
the results of the {\em ab initio} band structure calculations for
graphene \cite{Painter}.

The electron eigenfunctions of the total Hamiltonian can be
written in terms of $2p_z$ carbon orbitals and the adsorbed atom
EMOs. In this regard, it has been shown \cite{Robinson,George1}
that there is a relationship between $2p_z$ orbitals of graphene
and the adsorbate orbitals, so that, the total Hamiltonian Eq.
(\ref{01}) can be replaced by an effective Hamiltonian which
consists of the graphene $2p_z$ Hamiltonian with an
energy-dependent potential on each carbon atom to which the
adsorbed atom binds. In fact, the energy-dependent effective
potential acts like a self-energy that an adsorbate induces on
electrons in graphene.  For a single Co adatom that bonds to six
neighboring graphene carbon atoms, the effective Hamiltonian can
be written as \cite{George1}
\begin{equation}\label{1}
H_{\mathrm{eff}}=H_0+V
\end{equation}
\begin{equation}\label{2}
V=\sum_{n,m=1}^6\sum_{\sigma} V_{nm,\sigma}a^\dag_{n\sigma}
a_{m\sigma}
\end{equation}
where $a_{n\sigma}$ is the annihilation operator for an electron
in the 2$p_z$ orbital of carbon atom $n$ and
$V_{nm,\sigma}=\sum_\alpha\gamma_{\alpha
n,\sigma}\gamma^{*}_{\alpha
m,\sigma}/(\epsilon-\epsilon_{\alpha\sigma})$ displays a resonant
energy dependence. Here, the summation is over the extended
molecular orbitals $\alpha$ of the Co adatom that bonds to carbon
atoms $m$ and $n$ of the graphene. In this study, we consider
171~EMOs for the adsorbed Co atom on graphene which are linear
combinations of the Co $3d$, $4s$ and $4p$ valence orbitals and
the 2s, 2$p_x$, 2$p_y$ valence orbitals of 54 carbon atoms shown
in Fig. 1. The spin dependence of $\epsilon_{\alpha\sigma}$ and
$\gamma_{\alpha n,\sigma}$ originates from the spin splitting of
the Co $4s$ and $3d$ orbitals. Due to the crystal field splitting,
the Co $3d$ orbitals split into three symmetry groups: the singly
degenerate $A_1$ group which only consists of 3$d_{3z^2-r^2}$
orbital, a doubly degenerate $E_1$ group consisting of the
3$d_{zx}$ and 3$d_{zy}$ orbitals, and a doubly degenerate $E_2$
group derived from the 3$d_{xy}$ and 3$d_{x^2-y^2}$ orbitals. The
spin splitting values appear as extra energies that modify the Co
$4s$ and $3d$ orbitals on-site energies obtained from the extended
H\"{u}ckel theory parameters. The value of these splittings are
calculated from each of two major peaks in spin density of states
spectrum for Co from Ref. \cite{Chan}. In addition, since in the
parameters obtained from the extended H\"{u}ckel theory the effect
of all carbon atoms (neighbors) is included in the calculations,
the Dirac point of graphene will shift somewhat (0.072$t$) from
zero energy to higher energy. To compensate for this effect,
the cobalt energy levels are also shifted in
energy by the same amount.

\section{T-matrix and Unperturbed Green's function}
To study the strength of scattering associated with a single Co
adatom on graphene we use the $T$-matrix formalism which is
defined as $G^{\mathrm{eff}}=G^0+G^0TG^0$ where
$G^{\mathrm{eff}}=(\epsilon+i\eta-H_\mathrm{eff})^{-1}$ is the
effective Green's function for the graphene in the presence of the
Co adatom, and $G^0=(\epsilon+i\eta-H_0)^{-1}$ is the unperturbed
Green's function (i.e. graphene Green's function in the absence of
Co adsorbate) for $\pi$-band electrons. The $T$-matrix is related
to $G^0$ and $V$ by $T=V(1-G^0V)^{-1}$ where all the operators are
6$\times$6 matrices with the basis of the graphene $p_z$ orbitals
of the carbon atoms to which the Co atom binds. To obtain the
$T$-matrix components we need the matrix elements of $G_0$. Since,
$\varepsilon_{v}(\mathbf{k})=-\varepsilon_{c}(\mathbf{k})$ the
matrix elements of $\langle
i|G^0(\epsilon)|j\rangle=G^0_{ij}(\epsilon)$ between two arbitrary
sites $i$ and $j$ of a graphene sheet can be written as
\begin{equation}\label{29}
G^0_{ij}(\epsilon)=\sum_\mathbf{k}\left[\frac{\langle
i|\mathbf{k},+\rangle\langle\mathbf{k},+|j\rangle}{\epsilon+i\eta-\varepsilon_c(\mathbf{k})}
+\frac{\langle
i|\mathbf{k},-\rangle\langle\mathbf{k},-|j\rangle}{\epsilon+i\eta+\varepsilon_c(\mathbf{k})}\right]
\end{equation}
where the sum is over the Brillouin zone, $\epsilon$ is the energy
and $\eta$ is a positive infinitesimal. The results for
$G^0_{ij}(\epsilon)$ depend on whether the sites $i$ and $j$
correspond to two equivalent or inequivalent sites. In the case of
equivalent sites, the diagonal and off-diagonal matrix elements
can be obtained from the following equation
\begin{equation}\label{A36}
G^0_{ij}(\epsilon)=\frac{1}{N}\sum_\mathbf{k}
\left[\frac{\epsilon}{(\epsilon+i\eta)^2-\varepsilon^2_c(\mathbf{k})}\right]
e^{i\mathbf{k}\cdot(\mathbf{R}_{i}-\mathbf{R}_{j})}
\end{equation}
where $N=N_{A(B)}$. For the case of inequivalent sites, the
off-diagonal matrix elements of $G^0$ corresponding to two
inequivalent atoms ($A$ and $B$ in the same cells or in different
cells) are written as:
\begin{equation}\label{A37}
G^0_{ij}(\epsilon)=\frac{1}{N}\sum_\mathbf{k}
\left[\frac{-\varepsilon_c(\mathbf{k})}{(\epsilon+i\eta)^2-\varepsilon^2_c(\mathbf{k})}\right]
e^{i[\mathbf{k}\cdot(\mathbf{R}_{i}-\mathbf{R}_{j})+\phi{(\mathbf{k})}]}
\end{equation}
We note that for the cases of adsorbates such as H and OH that
bond to only one carbon atom and also for O adatom which bonds to
two neighboring carbon atoms, by expanding the band structure of
graphene close to the Dirac points (at the K and K' points in the
Brillouin zone), one can linearize the dispersion relation and
obtain analytic expressions for the required matrix elements that
is valid in the energy range $|\epsilon|/t\leq 0.8$ (see Ref.
\cite{George1}). For a general matrix element of the Green's
function, however, there is no such analytic expression, and we
therefore consider the full $\pi$-band structure of graphene and
evaluate numerically the above Green's function matrix elements
which are valid in all energy ranges.

The locations of resonance energies are determined by the poles of
$T$-matrix, which can be different for the spin-up and spin-down
states. By taking matrix elements of the effective Green's
function between the graphene 2$p_z$ orbitals of the six carbon
atoms to which the Co atom binds, we obtain
\begin{equation}\label{G}
G^{\mathrm{eff}}_{ij,\sigma}(\epsilon)=G^0_{ij}(\epsilon)
+\sum_{ll'=1}^6G^0_{il}(\epsilon)T_{ll',\sigma}(\epsilon)G^0_{l'j}(\epsilon)
\end{equation}
where $T_{ll',\sigma}(\epsilon)$ are the spin- and
energy-dependent $T$-matrix elements which describe resonant
scattering of electrons in graphene due to the adsorbate levels
renormalized by hybridization with $2p_z$ orbital of carbon atoms.
The effect of this hybridization usually appears as a shift in the
resonance levels towards the Dirac point of graphene
\cite{Robinson,George1}.

\section{Spin-dependent Dirac point resonance of Co adatom}
\subsection{$T$-matrix feature}
To investigate the scattering effects associated with a single Co
adatom on graphene, we study the square modulus of the $T$-matrix.
The scattering will be a spin-dependent process due to the
spin-splitting of Co 4$s$ and 3$d$ orbitals, and the scattering at
the energies $\epsilon$ at which $|T_{\sigma}(\epsilon)|^2$ has
maxima is of the  resonant type. The energies of these resonant
states can be found by setting the denominator of the square
modulus of the $T$-matrix equal to zero, i.e.
$|\mathrm{det}(1-G^0V)|^2=0$. Based on the $\gamma_{\alpha
n,\sigma}$ and $\epsilon_{\alpha n,\sigma}$ parameters obtained
from the present model, the square modulus of the $T$-matrix for
spin-up and spin-down electrons is shown vs. the electron energy
$\epsilon$ in Fig. 2. The two studied models (\textit{solid} and
\textit{dashed} curves) are explained in detail below. Note that,
the effect of overlaps $s_{\alpha n}$ between the EMOs
$\psi_{\alpha}$ and the 2$p_z$ orbitals $\phi_{n}$ of carbon atoms
$n$ are included in all of the calculations.
\begin{figure}
\centerline{\includegraphics[width=1.0\linewidth]{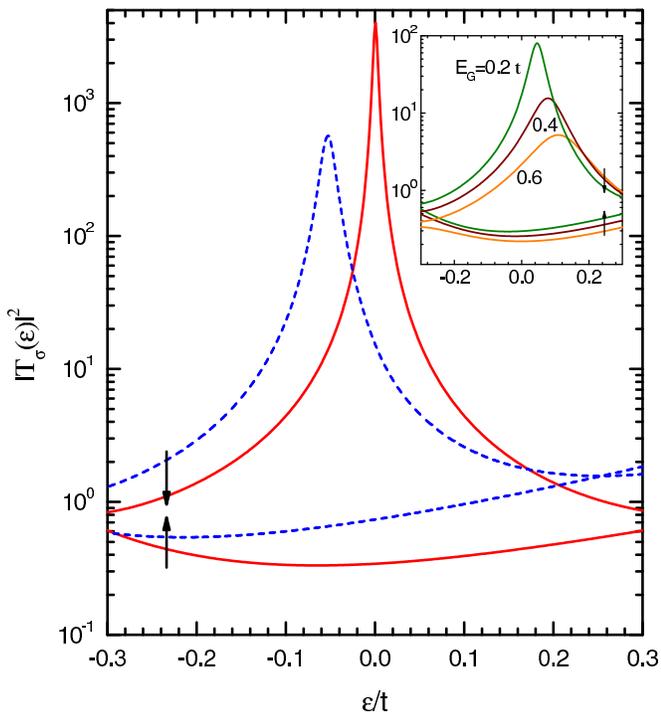}}
\caption{(Color online) Spin-dependence of square modulus of
$T$-matrix vs. electron energy at the gate voltage for which
$E_G$, the gate induced shift in energy of the graphene orbitals
relative to those of the Co, is zero. Solid lines correspond to a
case in which the EMOs are linear combinations of the atomic
valence orbitals of the adsorbed Co atom and the $2s$, $2p_x$, and
$2p_y$ valence orbitals of each of the C atoms, while the dashed
lines are related to a case in which only the valence orbitals of
the adsorbate (no carbon orbitals) are included in the EMOs. The
symbol $\uparrow$ ($\downarrow$) corresponds to spin-up
(spin-down) electrons. The inset shows the effect of gate voltage
on the square modulus of $T$-matrix when the $2s$, $2p_x$, and
$2p_y$ orbitals of carbon atoms are included in the calculations.
$T_\sigma$, $\epsilon$ and $E_G$ are in units of $t$=2.7 eV.}
\end{figure}

The solid curves correspond to a model in which all 171~EMOs are
used in the calculations. In other words, the EMOs are linear
combinations of the $4s$, $4p_x$, $4p_y$, $4p_z$, $3d_{x^2-y^2}$,
$3d_{xy}$, $3d_{3z^2-r^2}$, $3d_{zx}$, $3d_{zy}$ atomic valence
orbitals of the adsorbed Co atom and the $2s$, $2p_x$, and $2p_y$
valence orbitals of all 54 carbon atoms. The square modulus of the
$T$-matrix of the spin-down electrons displays a resonant peak at
the Dirac point of graphene, while for the spin-up electrons there
is no such a resonance feature in the shown energy interval
($|\epsilon|/t\leq 0.3$). This is a measure of high spin
polarization in the system.

The dashed curves correspond to a simpler model of the EMOs in
which no carbon orbitals are taken into account in the EMO
calculation and only 9 orbitals due to the Co atom are included in
the EMOs. We see that, the omission of the graphene carbon $2s$,
$2p_x$ and $2p_y$ valence orbitals in the calculations and of the
coupling of adsorbate to these orbitals (which are involved in the
local rehybridization of the graphene from $sp^2$ to $sp^3$
bonding near the adsorbed Co atom) shifts the position of the
adsorbate-induced Dirac point resonance energy to
$\epsilon_{DR}=-0.053 t$ and its intensity is about one order of
magnitude weaker than that found in the more complete model.

In earlier work \cite{Wehling3,George1,Sofo1,Sofo2} it has been
shown that the adsorption of H, F, OH, and O atoms on graphene
distorts the graphene plane by shifting carbon atoms out of the
plane. The omission of the graphene carbon $2s$, $2p_x$, and
$2p_y$ valence orbitals from EMOs was found to affect the
adsorbate-induced resonances for these adsorbates very strongly
\cite{George1}. On the other hand, we have already explained that
there is no out of plane structural distortion due to the Co
adsorption on graphene, while if these valence orbitals are
omitted from the EMOs $\psi_\alpha$, the Co resonance state is
still affected significantly. Since the inclusion of $2s$, $2p_x$,
and $2p_y$ valence orbitals in the EMOs plays a dominant role for
the Co adsorbate on graphene, and these orbitals play a crucial
role in the rehybridization of graphene from $sp^2$ to $sp^3$
bonding, it seems reasonable to infer that significant
rehybridization from the $sp^2$ to the $sp^3$ electronic structure
occurs for this system, despite the absence of structural out of
plane distortion of the graphene.

\begin{table*}[ht]\caption{Minimal set of effective tight-binding
parameters $\epsilon_{\alpha\sigma}$, $\gamma_{\alpha n,\sigma}$
and $s_{\alpha n}$ in units of $t$ = 2.7 eV for adsorbed Co
atom on graphene at zero gate voltage. The spin-dependent EMO
energies $\epsilon_{\alpha\sigma}$ are measured from the Dirac
point energy of graphene.}\label {T1}\centering\small
\begin{tabular}{ccccccccccccc}
\hline\hline
$~~~\epsilon_{\alpha\uparrow}~~$&$~~~\gamma_{\alpha1\uparrow}~~$&
$~~~\gamma_{\alpha2\uparrow}~~$&$~~~\gamma_{\alpha3\uparrow}~~$&
$~~~\gamma_{\alpha4\uparrow}~~$&$~~~\gamma_{\alpha5\uparrow}~~$&
$~~~\gamma_{\alpha6\uparrow}~~$&$~~~s_{\alpha1\uparrow}~~$&
$~~~s_{\alpha2\uparrow}~~$&$~~~s_{\alpha3\uparrow}~~$&
$~~~s_{\alpha4\uparrow}~~$&$~~~s_{\alpha5\uparrow}~~$&
$~~~s_{\alpha6\uparrow}~~$\\
\hline\centering
-0.855 &  -0.084  &  -0.078  &  ~0.162  &  -0.078  &  -0.079  &  ~0.157  &  ~0.023  &  ~0.022  &  -0.044  &  ~0.022  &  ~0.022  &  -0.043 \\
-0.855 &  ~0.135  &  -0.142  &  -0.001  &  ~0.140  &  -0.135  &  ~0.003  &  -0.037  &  ~0.039  &  ~0.000  &  -0.038  &  ~0.037  &  -0.001 \\
-0.704 &  ~0.175  &  ~0.239  &  ~0.067  &  -0.173  &  -0.244  &  -0.068  &  -0.044  &  -0.061  &  -0.017  &  ~0.044  &  ~0.062  &  ~0.017 \\
-0.704 &  ~0.177  &  -0.062  &  -0.236  &  -0.180  &  ~0.061  &  ~0.244  &  -0.045  &  ~0.016  &  ~0.060  &  ~0.046  &  -0.015  &  -0.061 \\
~1.363 &  -0.296  &  -0.293  &  -0.292  &  -0.296  &  -0.298  &  -0.300  &  ~0.102  &  ~0.101  &  ~0.101  &  ~0.102  &  ~0.102  &  ~0.103 \\
~2.513 &  -0.043  &  -0.276  &  -0.246  &  ~0.017  &  ~0.258  &  ~0.228  &  ~0.025  &  ~0.172  &  ~0.153  &  -0.012  &  -0.163  &  -0.144 \\
~2.513 &  -0.290  &  -0.117  &  ~0.171  &  ~0.295  &  ~0.125  &  -0.172  &  ~0.182  &  ~0.074  &  -0.107  &  -0.184  &  -0.078  &  ~0.108 \\
~3.201 &  ~0.161  &  ~0.158  &  ~0.157  &  ~0.162  &  ~0.165  &  ~0.166  &  -0.087  &  -0.086  &  -0.085  &  -0.087  &  -0.088  &  -0.089 \\
~4.291 &  ~0.391  &  ~0.380  &  ~0.377  &  ~0.390  &  ~0.402  &  ~0.405  &  -0.200  &  -0.194  &  -0.192  &  -0.200  &  -0.206  &  -0.207 \\
~8.174 &  -0.144  &  -0.161  &  -0.162  &  -0.148  &  -0.130  &  -0.129  &  ~0.071  &  ~0.074  &  ~0.074  &  ~0.072  &  ~0.068  &  ~0.068 \\
\hline \hline
\end{tabular}

\begin{tabular}{ccccccccccccc}
$~~~\epsilon_{\alpha\downarrow}~~$&$~~~\gamma_{\alpha1\downarrow}~~$&
$~~~\gamma_{\alpha2\downarrow}~~$&$~~~\gamma_{\alpha3\downarrow}~~$&
$~~~\gamma_{\alpha4\downarrow}~~$&$~~~\gamma_{\alpha5\downarrow}~~$&
$~~~\gamma_{\alpha6\downarrow}~~$&$~~~s_{\alpha1\downarrow}~~$&
$~~~s_{\alpha2\downarrow}~~$&$~~~s_{\alpha3\downarrow}~~$&
$~~~s_{\alpha4\downarrow}~~$&$~~~s_{\alpha5\downarrow}~~$&
$~~~s_{\alpha6\downarrow}~~$\\
\hline
-0.494   & ~0.084  &  ~0.088  &  -0.172  & ~0.074  &  ~0.090  &  -0.163  &  -0.022  &  -0.023  &  ~0.045  &  -0.019  &  -0.023  & ~0.042 \\
-0.494   & ~0.144  &  -0.148  &  -0.009  & ~0.151  &  -0.137  &  -0.003  &  -0.038  &  ~0.038  &  ~0.002  &  -0.039  &  ~0.036  & ~0.001 \\
-0.121   & ~0.128  &  ~0.263  &  ~0.140  & -0.121  &  -0.271  &  -0.144  &  -0.029  &  -0.060  &  -0.032  &  ~0.028  &  ~0.062  & ~0.033 \\
-0.121   & -0.234  &  -0.008  &  ~0.220  & ~0.238  &  ~0.013  &  -0.232  &  ~0.054  &  ~0.002  &  -0.050  &  -0.054  &  -0.003  & ~0.053 \\
~1.518   & ~0.263  &  ~0.262  &  ~0.261  & ~0.264  &  ~0.265  &  ~0.266  &  -0.092  &  -0.092  &  -0.092  &  -0.092  &  -0.092  & -0.093 \\
~2.514   & -0.042  &  -0.278  &  -0.248  & ~0.017  &  ~0.260  &  ~0.231  &  ~0.025  &  ~0.172  &  ~0.153  &  -0.012  &  -0.163  & -0.145 \\
~2.514   & ~0.293  &  ~0.118  &  -0.172  & -0.297  &  -0.126  &  ~0.173  &  -0.182  &  -0.074  &  ~0.107  &  ~0.185  &  ~0.078  & -0.108 \\
~3.210   & ~0.155  &  ~0.152  &  ~0.151  & ~0.155  &  ~0.159  &  ~0.160  &  -0.084  &  -0.083  &  -0.082  &  -0.084  &  -0.085  & -0.086 \\
~4.362   & ~0.395  &  ~0.382  &  ~0.379  & ~0.394  &  ~0.406  &  ~0.410  &  -0.204  &  -0.198  &  -0.196  &  -0.204  &  -0.211  & -0.212 \\
~8.186   & ~0.146  &  ~0.163  &  ~0.164  & ~0.149  &  ~0.132  &  ~0.131  &  -0.073  &  -0.076  &  -0.076  &  -0.074  &  -0.071  & -0.071 \\
\hline \hline
\end{tabular}
\end{table*}

In the experiment of Brar {\em et al.} \cite{Brar} both a back
gate and the STM tip played the role of gates. Together they
induced electrostatic potential differences between the graphene
plane and the adsorbed Co atom that manifested themselves in the
experimental STM data \cite{Brar}. We model this gate-induced
potential difference phenomenologically by shifting the carbon
orbitals energies, derived from the extended H\"{u}ckel theory,
and also the electronic structure of graphene $\pi$-band used in
the unperturbed Green's function of the graphene plane, $G^0$, by
a gate voltage-dependent amount $E_G$ relative to the energies of
the Co orbitals. Thus $E_G$ acts as a proxy for the gate voltage
in the present work. Accordingly, $E_G=0$ corresponds to the gate
voltage at which the graphene Dirac point and the resonant state
due to the adsorbate are located at zero energy $\epsilon/t=0.0$.
The gate voltage at which $E_G=0$ is an important reference point
for the gate-controlled ionization of the Co atom, especially when
we compare our differential conductance results with the STM and
spectroscopy measurements in the Sec. V. We emphasize that, in our
calculations, the extended H\"{u}ckel model parameters describing
both the on-site and intersite Hamiltonian matrix elements are
recalculated for each value of the gate voltage. The dependence of
the intersite Hamiltonian matrix elements on the gate
voltage\cite{shift} implies that hybridization between the
orbitals on different atoms depends on the gate voltage in the
present theory. We note that previous work of Sofo {\em et al.}
\cite {Sofo1,Sofo2} has indicated that doping can lead to a strong
change in hybridization of the graphene.

The application of a gate voltage in the experiment \cite{Brar} caused the Co density of
states features to move relative to the  Fermi energy and hence,
the Co levels could be emptied or filled with electron
carriers.\cite{Brar} The movement of a prominent peak (marked S in
the STM conductance spectra shown in Figs. 2a and 2b in Ref.
\cite{Brar}) relative to the Dirac point with changing gate
voltage can be interpreted as further evidence of a Co energy
level moving relative to the graphene $\pi$ band as the gate
voltage was varied. The strength of the conductance peak S was
observed to increase greatly in the experiment (Fig. 2a in Ref.
\cite{Brar}) as the peak approached the graphene Dirac point (Fig.
2b in Ref. \cite{Brar}). As will be discussed in Section V, we
interpret this behavior as being due to a Co resonant level
approaching the graphene Dirac point  as $E_G$ approaches zero. As
this happens the cobalt-induced resonance feature for spin-down
graphene electrons (associated with the peak in the square modulus
of $T$-matrix in Fig.2) also approaches at the Dirac point as
$E_G$ strengthens. However with increasing $|E_G|$, the
probability for overlap between the resonant cobalt  level for
spin-down electrons and the graphene states in the vicinity of the
Dirac point decreases and the sharp conductance peak is suppressed
in a similar way to the rather broad peak (marked S) in Fig. 2 in
Ref. \cite{Brar} for the higher absolute values of the gate
voltage.

The inset of Fig. 2 shows the calculated $E_G$ dependence of the
square modulus of the $T$-matrix and indicates that the electron
energy $\epsilon_{DR}$ at which the resonance occurs and the
strength of electron scattering depends on the value of gate
voltage. We have mentioned above that a sharp resonance peak
centered at $\epsilon_{DR}=0.0\,t$ appears at the gate voltage for
which $E_G=0$ due to spin-down electrons. However, with increasing
$E_G$ the strength of resonance peak in the inset of Fig. 2
decreases and the peak position shifts towards the higher
energies; $\epsilon_{DR}=0.045\,t, 0.078\,t, 0.11\,t$ at
$E_G=0.2\,t, 0.4\,t, 0.6\,t$ respectively. In the case of
decreasing $E_G$ the peak position $\epsilon_{DR}$, shifts to
lower energies. For spin-up electrons the variation of the
$|T_{\sigma}(\epsilon)|^2$ function vs. energy is approximately
the same in the all gate voltages. The reason of such a behavior
can be understood in terms of LDOS of the Co adsorbate and
graphene sheet that will be explained below.

\begin{figure}
\centerline{\includegraphics[width=1.0\linewidth]{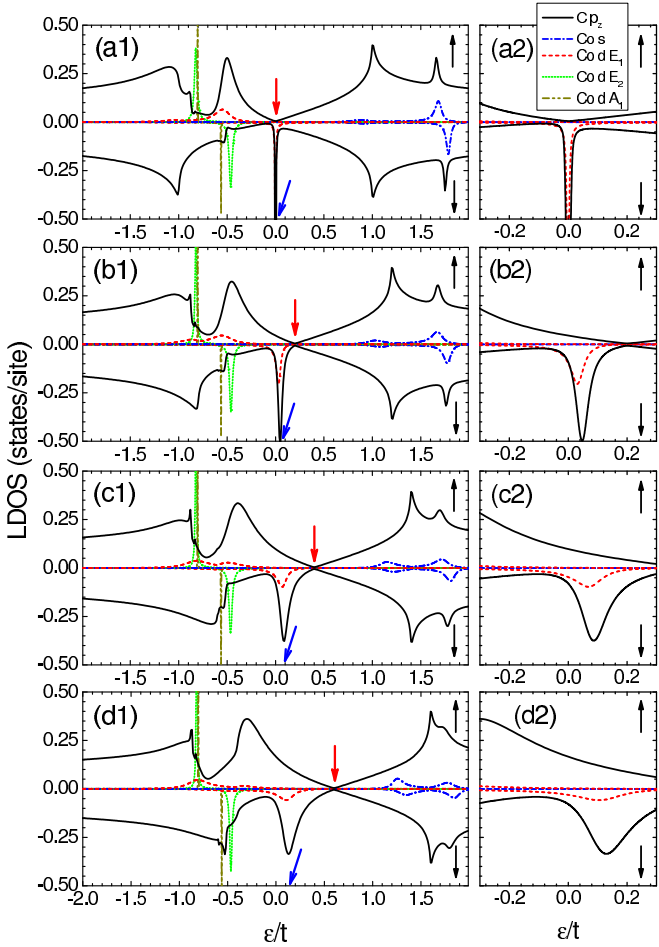}}
\caption{(Color online) Calculated LDOS per site vs. electron
energy $\epsilon$ for a single Co atom adsorbed on graphene at
various gate voltages: (a1), (a2) $E_G=0.0\,t$; (b1), (b2) $
E_G=0.2\,t$, (c1), (c2) $ E_G=0.4\,t$; (d1), (d2) $ E_G=0.6\,t$.
The symbol $\uparrow$ ($\downarrow$) corresponds to spin-up
(spin-down) electrons. The LDOS associated with the $2p_z$ orbital
of a graphene carbon atom to which the Co atom bonds (solid
black), Co $4s$ (dash-dotted blue), Co $3d$ $A_1$ (dash-dot-dotted
dark yellow), Co $3d$ $E_1$ (dashed red), and Co $3d$ $E_2$
(dotted green) states are shown. In Co LDOS, we multiplied by the
scale factor 1/50. The EMOs included in graphene LDOS calculation
are linear combinations of the atomic valence orbitals of the Co
atom and the $2s$, $2p_x$, and $2p_y$ valence orbitals of C atoms.
For Co LDOS the effect of carbon $2p_z$ orbitals  is included as a
self-energy in the calculations. The red and blue arrows indicate
the position of the graphene Dirac point and the graphene resonant
state, respectively.}
\end{figure}

The EMO energies $\epsilon_{\alpha\sigma}$, hopping
$\gamma_{\alpha n,\sigma}$ and overlap $s_{\alpha n}$
parameters for a minimal tight-binding model Hamiltonian that
describes the Dirac point resonant state involved in the
spin-dependent square modulus of the $T$-matrix and graphene LDOS
at zero gate voltage are presented in Table I.

\subsection{Local Density of States}
The calculated spin-dependent LDOS for graphene in the presence of
an adsorbed Co atom, and for the Co $4s$, $A_1$, $E_1$, and $E_2$
orbitals at different gate voltages are shown in Fig. 3. For
greater clarity, the electronic states within an energy window
close to the Dirac point of graphene are also shown separately
[see Fig. 3(a2)-(d2)]. In the case of the graphene LDOS, we used
the same extended H\"{u}ckel-theory-based model in the
calculations that was used to calculate the square modulus of the
$T$-matrix. Thus, the model includes the effects of the local
rehybridization of the graphene plane due to the presence of the
Co adatom. The LDOS in this case is defined as
$D_{i,\sigma}(\epsilon)=-\mathrm{Im}[\langle
i|G^{\mathrm{eff}}_{\sigma}(\epsilon)|i\rangle]/\pi$, where
$G^{\mathrm{eff}}_{\sigma}(\epsilon)$ is given by Eq.(\ref{G}) and
$|i\rangle$ represents a 2$p_z$ orbital of a carbon atom. The
solid curves in Fig. 3 correspond to the $2p_z$ orbital of a
graphene carbon atom to which the Co atom bonds. On the other
hand, for the LDOS of the Co adsorbate, we have used the
Hamiltonian
$\widetilde{H}_{\sigma}=H_{\sigma}^{\mathrm{Co}}+\Sigma^\mathrm{C}_\sigma(\epsilon)$
for an electron with spin $\sigma$ ($=\uparrow$ or $\downarrow$)
where $H_{\sigma}^{\mathrm{Co}}$ is the Hamiltonian of a free Co
atom whose matrix elements are given by the extended H\"{u}ckel
theory, and the self-energy
$\Sigma^\mathrm{C}_\sigma(\epsilon)=\tau_{\mathrm{Co-C},\sigma}
G^0(\epsilon)\tau^\dag_{\mathrm{Co-C},\sigma}$ contains the
coupling effects of the adsorbate to the graphene carbon atoms
\cite{Datta}. Here, $\tau_{\mathrm{Co-C},\sigma}$ is a hopping
matrix between the cobalt atom and the $2p_z$ orbitals of the six
carbon atoms of the graphene whose matrix elements are also given
by the extended H\"{u}ckel model parameters and $G^0(\epsilon)$ is
the Green's function of the clean graphene which is given by Eqs.
(\ref{A36}) and (\ref{A37}). In the above self-energy, the effect
of $2s$, $2p_x$, and $2p_y$ valence orbitals of C atoms could also
be considered in the calculation. In such a case, however, the
$\pi$-band Green's function, $G^0(\epsilon)$, is no longer
sufficient and the graphene $\sigma$-band Green's function is
required. Since the present self-energy is still able to reproduce
the expected resonance features, we consider just the hopping
parameters between the cobalt orbitals and the graphene $2p_z$
orbitals in the Co LDOS calculation. Note that the overlaps
$s_{\mathrm{Co-C}}$ between the Co atom and the $2p_z$ orbitals of
the carbon atoms to which the adsorbate bonds are included in the
calculations. Accordingly, the LDOS for the adsorbate is
determined by
$\widetilde{D}_{\ell,\sigma}(\epsilon)=-\mathrm{Im}[\langle
\ell|\widetilde{G}_{\sigma}(\epsilon)|\ell\rangle]/\pi$, where
$|\ell\rangle$ represents the $4s$, $A_1$, $E_1$, or $E_2$
orbital, and $\widetilde{G}_{\sigma}(\epsilon)=
(\epsilon+i\eta-\widetilde{H}_{\sigma})^{-1}$ is the Green's
function of the Co atom adsorbed on the graphene plane.

From Fig. 3 it is clear that the spin-down graphene states show a
resonance at the energies close to the Dirac point that is due to
coupling between the graphene and the Co $E_1$ orbital. This
resonance is exactly positioned at the Dirac point energy at the
gate voltage for which $E_G$, the gate induced shift in energy of
the graphene orbitals relative to those of the Co, is zero. For
this gate voltage the spin-down Co $E_1$ state coincides with the
resonant graphene state that appears due to the presence of the Co
adatom. Furthermore, because of the spin-splitting of the Co $3d$
orbitals, the peak energy of the spin-up Co $E_1$ state is somewhat
far (i.e., at approximately $-0.55\,t$ for $E_G=0$) from the Dirac point
energy and accordingly, only a relatively broad peak due to the
spin-up Co $E_1$ state appears in the spectrum. The Co $A_1$ and
$E_2$ orbitals are completely occupied for both spin states at the
gate voltage for which $E_G=0$. Note that, the Fermi energy $E_F$
is set at the Dirac point of graphene, i.e. $E_F=0.0\,t$. In this
study, only the effect of positive $E_G$ is investigated. The
reason for this will be explained in the next section. In both the
graphene and Co LDOSs there is an energy shift with changing gate
voltage. Although, as we mentioned above, the energy level shift
due to the gate voltage is only applied on the graphene plane, the
Co orbitals are also affected by this effect because, due to the
non-zero overlap between the Co and graphene atomic orbitals, the
effective hopping energies between the carbon and the Co orbitals
depend on the on-site energies of the carbon orbitals
\cite{Emberly1,Emberly2}. In the energy window ($|\epsilon|/t\leq
0.3$) that is shown in Fig. 3, the gate voltage decreases the
occupation of both spin-up and spin-down graphene states by
shifting the electronic states above the Fermi level $E_F$, and
hence the occupation of spin-down $E_1$ orbital is also decreased.
This result can be clearly seen in Fig. 3 where $E_G$ increases
from 0.0$t$ to 0.2$t$ to 0.4$t$ and 0.6$t$ and the Dirac point of
the graphene (indicated by red arrows) moves from zero energy to
the values of 0.2$t$, 0.4$t$ and 0.6$t$ , respectively. Therefore,
the Co orbitals which are coupled to the graphene shift somewhat
to the higher energies and the Co $E_1$ orbital is gradually
emptied of electrons. This result for Co $3d$ orbitals is in
agreement with the recent DFT calculations for gated Co atom on
graphene in supercells with periodic boundary conditions
\cite{Chan,Jacob,Wehling8}. Furthermore, by comparing Figs.
3(a)-3(d) it is clear that, with increasing $E_G$, the amplitudes
of the Co $E_1$ and graphene $p_z$ density of states peaks
decrease, the energy difference between the Dirac point and the
graphene resonance state position increases and both the Co $E_1$
level and the graphene resonance state induced by Co adatom are
broadened.

This indicates that the Co $E_1$ orbital is coupled strongly to
the $\pi$-orbitals of graphene and a gate voltage can ionize the
Co adatom through this orbital. This is in agreement with the work
of Liu \textit{et al.} \cite{Liu} in which they reported that the
hybridization between $E_1$ orbital of the $3d$-transition-metal
adatoms on graphene and $p_z$ orbitals of the graphene carbon
atoms is strong and responsible for strong covalent bonding with
graphene. By contrast, both the $A_1$ and $E_2$ orbitals are
strongly localized and far from the Dirac point, however, the
$A_1$ orbital hybridizes more weakly than the $E_2$ orbital with
the graphene states. Note that, comparison of the work of Liu {\em
et al.} \cite{Liu} with the present results shows somewhat
different coupling strengths of the Co $E_2$ orbital with $2p_z$
orbitals of the carbon atoms which may be attributed to the
periodic nature of the structures ($4\times4$ Co-graphene
supercell) studied in their calculations. It may also be
relevant that density functional
theory-based electronic structure calculations are known
to have fundamental limitations that can strongly affect the results that
they yield for the coupling between adsorbed atoms and graphene;
see Section VII of Ref. \onlinecite{George1} for a discussion and specific examples.
In addition, the spin-up and spin-down Co 4$s$ orbitals are far from the $E_F$; thus the
occupation of these orbitals is zero at low gate voltages,
although these orbitals hybridize strongly with the graphene
states. This result is different than the results for periodic
structures \cite{Mao,Wehling5,Chan} in which the Co $4s$ orbital
is close to the Dirac point energy and the effect of gating can
ionize the Co atom by occupying or unoccupying the $4s$ orbitals.
In addition, in the periodic structures, the Dirac point
resonances are broadened due to the presence of multiple adsorbed
atoms on the graphene \cite{Mao,Wehling5,Chan}. Therefore, it is
evident that to better understand the features of
adsorbate-induced Dirac point resonances and to verify the
veracity of the theoretical predictions, comprehensive
experimental investigations would be very important.

\section{Simulation of STM measurements}
\subsection{Model}
In this section, assuming coherent transport, we describe how the
electronic differential conductance ($dI/dV$) through a Co atom
adsorbed on a graphene plane can be calculated using the
tight-binding Hamiltonians described in the preceding sections and
the Landauer-B\"{u}ttiker formalism \cite{Datta}. The aim of this
is to make a comparison with the STM differential conductance
measurements in which the STM tip was held directly above the
individual Co adatoms deposited onto a back-gated graphene device
at 4.2 K \cite{Brar}. To simulate the experimental condition, we
model the STM tip (shown in Fig. 1) by a one-dimensional chain of
Pt atoms with nearest neighbor hopping integral $t$. The Fermi
energy is fixed at the center of the Pt chain band structure. For
Pt atoms we use $6s$, $6p_x$, $6p_y$, $6p_z$, $5d_{x^2-y^2}$,
$5d_{xy}$, $5d_{3z^2-r^2}$, $5d_{zx}$, $5d_{zy}$ atomic valence
orbitals. Since each Pt atom in the tip is described by 9 valence
orbitals, then we model the tip by nine semi-infinite leads of
orthogonal atomic orbitals, one orbital per lead site, with
spacing (periodicity) $a$ between orbital sites. Each of the
one-dimensional leads is decoupled from the others, thus, there is
no hopping between different lead orbitals. According to the above
description, the Green's function of the Co atom coupled to the
graphene carbon atoms and the STM system can be written as
\begin{equation}\label{DG}
G_\sigma(\epsilon,V)=[\epsilon+i\eta-H_{\sigma}^{\mathrm{Co}}
-\Sigma^\mathrm{C}_\sigma(\epsilon)-\Sigma^{\mathrm{STM}}_\sigma(\epsilon+eV)]^{-1}\
,
\end{equation}
where
$\Sigma^{\mathrm{STM}}_\sigma(\epsilon)=\tau_{\mathrm{Co-STM},\sigma}
g_{\mathrm{STM}}(\epsilon)\tau^\dag_{\mathrm{Co-STM},\sigma}$ is a
self-energy for an electron with spin $\sigma$ and describes the
coupling of the adsorbate to the STM tip,
$\tau_{\mathrm{Co-STM},\sigma}$ is the hopping matrix between STM
and the Co adatom whose matrix elements are given by the extended
H\"{u}ckel theory parameters, and
$g_\mathrm{STM}=-(1/t)\,e^{ika}\mathbb{I}$ is the surface Green's
function matrix for the semi-infinite leads where $k$ is the wave
number in the leads and $\mathbb{I}$ is a $9\times 9$ unit matrix
\cite{Datta,saffar1}. The overlaps $s_\mathrm{Co-STM}$ between the
Co and Pt orbitals are included in the calculations. Now, if we
make use of the nonequilibrium Green's function technique to
obtain the spin current $I_\sigma$ for a sample bias $V$ between
graphene and the tip \cite{Datta,saffar1}, then the spin-dependent
conductance $dI_\sigma/dV$ for an electron with spin $\sigma$ is
given by
\begin{equation}\label{didv}
\mathcal{G}_\sigma(V)\simeq
\frac{e}{h}\int_{-\infty}^{\infty}T_\sigma(\epsilon,V)\left(-\frac{\partial
f_{\mathrm{STM}}(\epsilon,V)}{\partial V}\right)d\epsilon\  ,
\end{equation}
where $f_\mathrm{STM}(\epsilon)$ is the tip Fermi distribution
function,
$T_\sigma(\epsilon,V)=\mathrm{Tr}[\Gamma^\mathrm{C}_\sigma(\epsilon)G_\sigma(\epsilon,V)
\Gamma^{\mathrm{STM}}_\sigma(\epsilon+eV)G_\sigma^{\dagger}(\epsilon,V)]$
is the Landauer transmission function that can be calculated from
the device Green's function, Eq. (\ref{DG}), and the broadening
functions which can be expressed for electrode $\alpha$ (= C or
STM) as
$\Gamma^\alpha_{\sigma}=-2\mathrm{Im}(\Sigma^\alpha_{\sigma})$. In
Eq. \ref{didv}, the voltage dependence of transmission function
has been ignored, because the transmission values are not
sensitive to low value bias voltages.

\subsection{Results}
The differential conductance is the most direct way to measure the
energy-dependent LDOS of a system. In the present system, the
calculated differential conductance,
$\mathcal{G}(V)=\mathcal{G}_\uparrow(V)+\mathcal{G}_\downarrow(V)$
vs. sample bias $V$ for different applied gate voltages is shown
in Fig. 4. Note that, the sample bias, $V$, means the voltage of
the graphene with respect to the tip. In the present study, the
graphene remains at the zero potential and the tip is at negative
voltages, similarly to the experiment. Hence, when a negative
voltage is applied to the tip, the sample bias is positive ($V>0$)
\cite{Brar}. Furthermore, the gate-dependent $\mathcal{G}(V)$
spectra in the experiment, were measured at negative gate
voltages. The measurements showed that, when the gate voltage
moves to more negative values, the Dirac point position shifts to
positive energies [see Fig. 2b in Ref. \cite{Brar}]. This behavior
is clearly equivalent to application of positive $E_G$ values as
we have done in Section IV and can be seen in Fig. 3.

The calculated differential conductance $\mathcal{G}(V)$ shows a
sharp peak at $E_G=0$ that comes from the Co spin-down resonant
state. The height of this peak decreases while the spectrum is
broadened as the gate voltage is varied and $|E_G|$ increases.
This effect is due to the nature of resonant states induced in the
graphene near the Dirac point by the presence of an adsorbate; as
can be seen in Fig. 3 these resonant states become narrower and
stronger as the associated energy level of the Co adsorbate
approaches the Dirac point energy,  i.e., $E_G \rightarrow 0$.
When $|E_G|$ increases, the cobalt orbital goes out of resonance
with the graphene density of states peak in the vicinity of the
Dirac point and the sharp conductance peak is suppressed.

In the experiment of Brar {\em et al.} \cite{Brar}, it has been
demonstrated that the difference between the location of Dirac
point voltage (marked $\mathrm{V_D}$ in Fig. 2b of Ref.
\cite{Brar}), associated with the Dirac point energy \cite{Brar2},
and the peak in the differential conductance signal (marked S in
Fig. 2a and 2b of Ref. \onlinecite{Brar}) increases as the
magnitude of experimental gate voltage is increased. This
enhancement in the difference between the Dirac point energy and
the energy of the S state with increasing the gate voltage is in
good agreement with the result of Fig. 3 which shows that,
although with increasing $E_G$ both the cobalt-induced resonance
feature (blue arrows) and the Dirac point energy (red arrow) in
the LDOS spectrum of Fig. 3 move in the same direction, the Dirac
point location moves more rapidly and the difference in energy
between both locations (red and blue arrows) increases with $E_G$.

The experimental data \cite{Brar} exhibits a sharp resonant peak S
in the $\mathcal{G}(V)$ spectrum at low gate voltages in Fig. 2a
of Ref. \onlinecite{Brar}. With increasing absolute value of the
gate voltage, the height of the peak decreases and the peak moves
to lower sample bias. The sharp peak seen for $E_G=0$ in our model
(Fig.4), is similar to the observed resonant peak S in Fig. 2a of
Ref. \onlinecite{Brar} at low gate voltages and evolves with gate
voltage in a qualitatively similar way as well.

The peaks in the $\mathcal{G}(V)$ spectra in Fig.4 occur
at the sample bias voltages at which the STM tip Fermi energy
crosses the peaks of the partial densities of states that are due to the Co d $E_1$
state and the associated Co-induced graphene resonant state that are seen
in Fig.3 a2-d2. These density of states features in Fig.3 a2-d2
{\em and the peaks in $\mathcal{G}(V)$ spectra in Fig.4} are strongest and narrowest
when the energy of the Co d $E_1$ state coincides with the graphene  Dirac
point energy, which occurs for $E_G=0$. The same is true of the Dirac point
resonance peak of the $T$-matrix as can be seen in Fig.2 and its
inset. The large increase in the intensity of the experimentally
observed resonant peak S in Fig. 2a of Ref.
\onlinecite{Brar} (with decreasing absolute value of the gate
voltage) also occurs as the S-peak energy approaches that
of the graphene Dirac point energy (labelled $V_D$) in Fig. 2b of Ref.
\onlinecite{Brar}. We conclude that the large increase in the intensity of
the resonant peak S in Fig. 2a of Ref.
\onlinecite{Brar} with decreasing absolute value of the gate
voltage is the experimental STM signature of the strengthening and narrowing
of Dirac point resonance
induced in the graphene by the Co (and of the associated Co level as well)
as the Co level crosses the graphene Dirac point energy.
Thus we propose that STM is able to detect the presence of
adsorbate-induced Dirac point resonances in the graphene
by monitoring the intensity of conductance peaks associated with
orbitals of the adsorbate as those adsorbate levels cross the
graphene Dirac point with changing gate voltage. Systematic
experimental studies of adsorbate-induced Dirac point resonances
in graphene based on this finding would be of considerable interest
since experimental studies of these resonances have, to our knowledge, not been
reported previously despite the considerable theoretical interest\cite{Mao,Chan0,Wehling5,Wehling6,
Can,Jacob,Chan,Rappoport,Wehling8,Liu,Skrypnyk1,Pereira1,Wehling1,
Skrypnyk2,Robinson,Pereira2,Basko,
Wehling2,Wehling3,Pershoguba,Wehling4,Skrypnyk3,George1} in this topic.

The gate dependence of $\mathcal{G}(V)$ also indicates that the
value of $E_G$ determines the ionization state of the Co atom,
because by changing the value of $E_G$, the Co $3d$ states can
cross the Fermi energy and accordingly the charge occupation in
these states changes. Therefore, the application of gate voltage
enables tuning the charge-carrier density of graphene and hence
the variation of Co adatom ionization state.

In the present model, the distance between the tip and the Co atom
is $h$=4 {\AA}. Therefore, the overlap and hence the direct
coupling between the Co atom and the tip is weak and that between
the tip and the graphene is much weaker still. The reason is that,
the sum of Pt and Co atomic radii is $\thicksim 2.64$ {\AA} which
is considerably less than $h$ which results in a small overlap
between their atomic wave functions. In reality, the Co levels
couple the STM tip to the graphene to open a conductive channel
for electron transport. Since the coupling between the
$E_1$-levels of adsorbate and the STM tip is weak the transport
process for such levels is tunneling and the differential
conductance spectra are quite similar to the $E_1$ resonance state
LDOS seen in Fig. 3. From the extended H\"{u}ckel theory hopping
parameters between the Co atom and the Pt atom, we found that the
$s$-levels of both atoms hybridize more strongly than that the
other levels. The hopping energy between $s$-levels is roughly one
order of magnitude greater than the hopping energy between
$E_1$-levels. Note that, in this work because of the position of
the Co $E_1$-level relative to the graphene Dirac point and also
to the other levels such as Co $4s$-level, the $E_1$ level plays
the main role in the tunneling transport through the cobalt atom
at low values of the gate voltage.
\begin{figure}
\centerline{\includegraphics[width=1.0\linewidth]{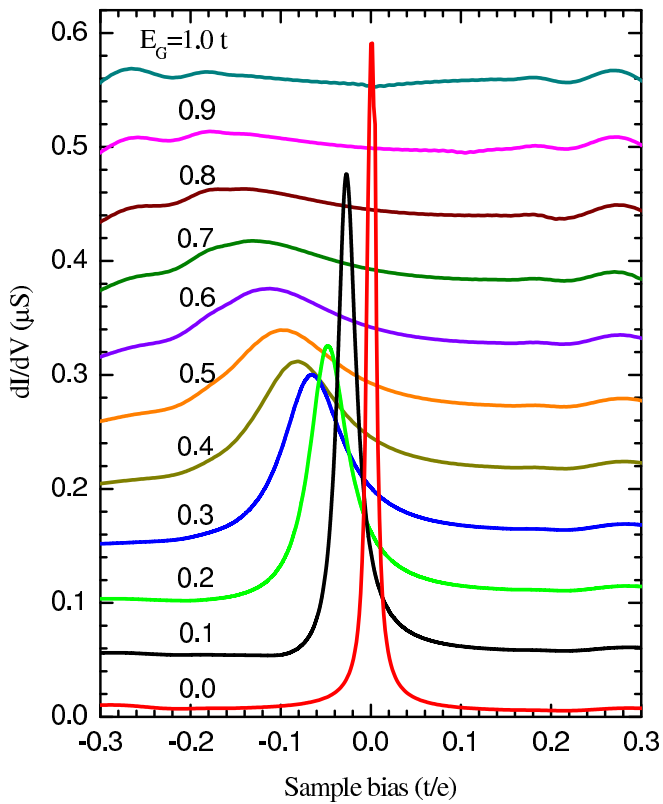}}
\caption{(Color online) Calculated total differential conductance
vs. sample bias for the graphene-cobalt-STM system, shown in
Fig.1, at various $E_G$ values. The curves are shifted by offsets
of 0.05 $\mu S$ with respect to each other.}
\end{figure}

Finally, it should be mentioned that we also examined the electron
transmission through the Co adatom in the graphene-Co-Pt structure
using the Landauer theory and the Lippmann-Schwinger equation
\cite{George3} in which the EMOs included in the graphene were
linear combinations of the atomic valence orbitals of the Co atom
and the $2s$, $2p_x$, and $2p_y$ valence orbitals of graphene
carbon atoms. The results, except with a small shift in the energy
of the resonant states, were the same that we obtained based on
the present method. This confirms that by the inclusion of just
carbon $2p_z$ orbitals in the graphene self-energy, we can obtain
the same resonance feature in the conductance spectrum that we
obtained based on the Lippmann-Schwinger approach.

\section{Discussion}

In this study, we have presented a tight-binding theory of the
Dirac point resonances due to a single Co adatom on gated graphene
based on the $\pi$-band tight-binding description of the graphene
plane with nearest neighbor hopping energy $t$ and the extended
H\"{u}ckel model of the adsorbate and the adsorbate-induced local
$sp^3$ rehybridization of the graphene. Using the generalized
effective Hamiltonian \cite{George1} for the relaxed geometry of
Co-graphene system based on $ab$ $initio$ density functional
theory, we found a strong spin-dependent scattering resonance at
the Dirac point of graphene and a gate dependence of this
resonance. This resonance arises from a hybridization of Co atomic
orbitals with graphene states and such resonances have been
predicted for $3d$ transition metals adsorbed on graphene
\cite{Wehling5,Mao,Jacob,Can,Chan}. The energy position of the Co
atomic levels with respect to the graphene Fermi energy determines
the ionization state of the adsorbate. Because of spin-splitting
of Co $4s$ and $3d$ orbitals, this ionization is different for
spin-up and spin-down states. The results showed that, the Co
$E_1$ state can be emptied with electrons as the state shifts
above the Fermi energy through application of gate voltage, in
agreement with the recent experiment \cite{Brar}. The change in
the occupation of the Co $3d$ orbitals and hence in the ionization
state of the adatom by application of gate voltage has also been
predicted using DFT-based calculations for periodic systems.
\cite{Jacob,Chan}.

Furthermore, we presented a formalism that we used to model the
differential conductance measurements. Our calculation shows that,
when the Co adatom states cross the Dirac point, a sharp feature
is induced in the $\mathcal{G}(V)$ spectrum which is consistent
with experiment and has not been addressed theoretically,
previously. The height, broadening and the position of this
feature is strongly dependent on gate voltage. The agreement
between this theoretical dependence and the experimentally
observed behavior of the S-peak in the STM differential
conductance spectrum \cite{Brar} allows us to attribute the S-peak
to tunneling between the STM tip and a cobalt-induced Dirac point
resonant state of the graphene, via a Co $3d$ orbital.
Systematic
experimental studies of adsorbate-induced Dirac point resonances
in graphene based on this finding would be of interest.

Our results also show the Dirac point energy of graphene and the
cobalt-induced resonance feature in the differential conductance
spectrum to move in different ways when a gate voltage is applied.
Therefore the ionization state of Co adatom can be tuned by
application of a gate voltage which induces a change in the
occupation of the Co $3d$ states. From the above findings
regarding the $T$-matrix and the electronic density of states  we
can conclude that the Co adatom on graphene has a strongly
spin-polarized LDOS especially at electron energies in the
vicinity of the Dirac point of graphene. This suggests that Co
adatoms on gated graphene may have possible applications in
magnetic data storage and magnetic sensors and other applications
which may be discovered.

\section*{ACKNOWLEDGEMENTS}
We thank M. Crommie for helpful correspondence.
This work was supported by NSERC, CIFAR, WestGrid, and Compute
Canada.

\end{document}